\documentclass[12pt,a4paper]{article}
\usepackage[utf8]{inputenc}
\usepackage[T1]{fontenc}
\usepackage[english]{babel}

\usepackage{graphicx} 
\usepackage[textfont=bf, font = normalsize]{caption} 
\usepackage{floatrow} 
\usepackage{amssymb}
\usepackage{amsfonts}
\usepackage{amsmath}
\usepackage{amssymb}
\usepackage{amsmath}
\usepackage{amsfonts}
\usepackage{makeidx}
\usepackage{epstopdf}
\usepackage{color}
\usepackage{breqn}
\usepackage[labelfont=bf]{caption}
\usepackage{float}
\usepackage{multirow}
\usepackage{multicol}
\usepackage{breqn}
\usepackage[colorlinks = true]{hyperref}
\floatstyle{plaintop} 
\restylefloat{table} 
\usepackage{subfigure} 
\usepackage[width=18.00cm, height=26.00cm]{geometry}
\title{\textbf{Understanding the training of PINNs for unsteady flow past a plunging foil through the lens of input subdomain level loss function gradients}}
\author{\textbf{Rahul Sundar\textsuperscript{1*}, Didier Lucor\textsuperscript{2}, and Sunetra Sarkar\textsuperscript{1}}\\\\
\textsuperscript{1*}\small{Department of Aerospace Engineering, Indian Institute of Technology Madras, Chennai 600036, India}\\
\textsuperscript{2} \small{Laboratoire Interdisciplinaire des Sciences du Numérique LISN-CNRS,}\\\small{ Universit\'e Paris-Saclay, Orsay 91403, France}\\$^*$\small{Email: rahulsundar@smail.iitm.ac.in}}
\date{}
\begin{document}
\maketitle
\begin{abstract}

Recently immersed boundary method-inspired physics-informed neural networks (PINNs) including the moving boundary-enabled PINNs (MB-PINNs) have shown the ability to accurately reconstruct velocity and recover pressure as a hidden variable for unsteady flow past moving bodies. Considering flow past a plunging foil, MB-PINNs were trained with global physics loss relaxation and also in conjunction with a physics-based undersampling method, obtaining good accuracy. The purpose of this study was to investigate which input spatial subdomain contributes to the training under the effect of physics loss relaxation and physics-based undersampling. In the context of MB-PINNs training, three spatial zones: the moving body, wake, and outer zones were defined. To quantify which spatial zone drives the training, two novel metrics are computed from the zonal loss component gradient statistics and the proportion of sample points in each zone. Results confirm that the learning indeed depends on the combined effect of the zonal loss component gradients and the proportion of points in each zone. Moreover, the dominant input zones are also the ones that have the strongest solution gradients in some sense. 
\end{abstract}

\section{{\textbf{INTRODUCTION}}}
Physics-informed neural networks have become promising for their use in solving complex inverse problems such as hidden physics recovery, data-driven equations discovery,  uncertainty quantification, 
Plain vanilla PINNs (\cite{raissi2019physics} are however difficult to train when the systems exhibit strong spatiotemporal gradients. Hence, recently, authors have proposed different strategies to train PINNs better, such as adaptive sampling, modified architectures, static and dynamic loss weighting.

Surrogate modeling of unsteady flow past moving boundaries such as flapping wings become challenging if the underlying high-fidelity simulation data used for such an endeavour is obtained using the immersed boundary method~~\cite{peskin2002immersed, balajewicz2014reduction}. This is because, IBM uses a fixed Eulerian background grid for the fluid, whereas the solid boundary is described by a set of Lagrangian markers. At any time instant, there exist eulerian grid cells bounded by the solid boundary which consists of fictitious flow field data. 

Flows past flapping wings are often characterized by strong flow-field gradients which makes it challenging for PINNs to train. Hence, in a recent work, a moving boundary enabled PINNs (MB-PINNs)~\cite{sundar2023physicsinformed} inspired by the IBM was proposed, where a fixed Eulerian grid for the fluid domain was considered discarding the solid region points a any given time and the no-slip boundary condition was enforced directly on the Lagrangian markers unlike in discrete forcing IBM~\cite{majumdar2020capturing}. A combination of global physics loss relaxation and vorticity cutoff based under sampling was shown to improve data efficiency while maintaining good accuracy in velocity reconstruction and simultaneous pressure recovery as a hidden variable. 

Since loss component weighting improves the loss balancing globally as reported in ~\cite{wang2020understanding}, it would be worthwhile to investigate if there exist local imbalances in the spatial domains of interest. This would also in a way determine which spatial region drives the training. One way of understanding how PINNs train is to look into the layer-wise loss component gradients as in \cite{wang2020understanding}  to understand the effect of competing objectives on the training. While these gradients are often visualized for an entire full batch setting, it is not visualized however in a localized context to understand the contributions from respective spatiotemporal zones. Visualizing the layer-wise gradients obtained for the loss components over specific spatiotemporal zones would highlight any imbalances in the gradient updates and also validate the need for localized weighting strategies further like in the self-adaptive weighting or residual-based attention approaches~\cite{xiang2022self, anagnostopoulos2023residual}. 

Hence, the current study's objectives are to investigate how loss contributions spatially are affected by physics loss relaxation combined with a physics-based undersampling and devise metrics to quantify which spatial zone drives the training of the network. As an example, the flow past a plunging airfoil at a low Reynolds number is considered following ~\cite{sundar2023physicsinformed}. From \cite{sundar2023physicsinformed}, three MB-PINN cases are considered with one being a baseline without any loss weighting or under-sampling and the other two with loss relaxation and undersampling.

The outline of the paper is as follows. The methodology of MB-PINNs and understanding the loss component gradients is presented in section~\ref{sec2}. The numerical experiments and results are discussed in section \ref{sec3} and finally the conclusions are presented in section~\ref{sec4}.

\section{\textbf{METHODOLOGY}}\label{sec2}
In this section, the problem setup, MB-PINN formulation, zonal splitting of loss component gradients, and metrics to quantify the zonal imbalances and determine the spatial zone driving the training will be discussed. 
\subsection{\textbf{The plunging foil system}}
\begin{figure}
	\centering
	\includegraphics[width=0.8\linewidth]{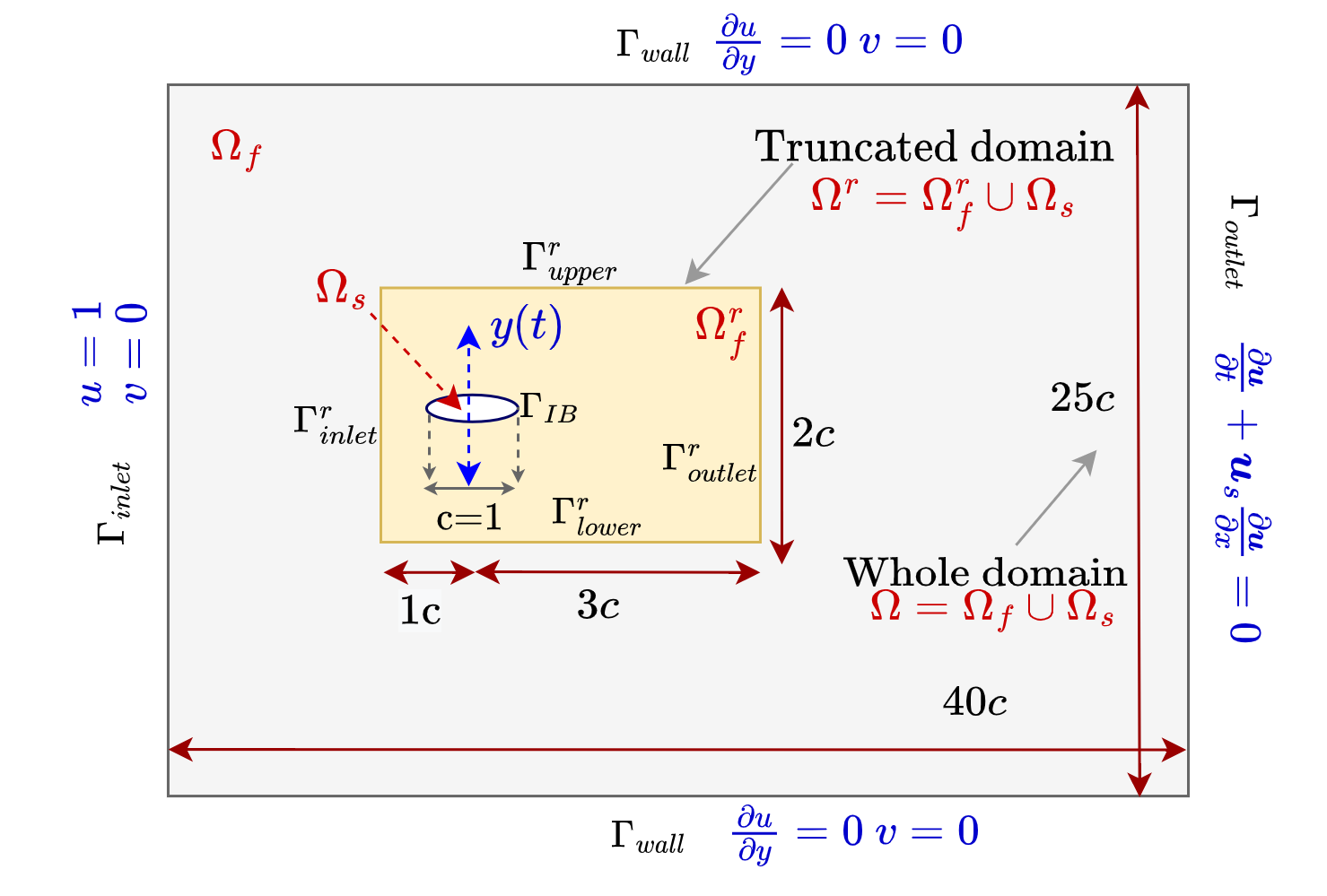}
	\caption{Problem setup (Image adapted from Sundar {\it et al.)}~\cite{sundar2023physicsinformed}.}
	\label{fig:problemsetup}
\end{figure}

In the present study, incompressible unsteady flow past a harmonically plunging elliptic foil is considered as an example. The flow around the foil is governed by the two-dimensional incompressible Navier-Stokes equations
\begin{align}
    \partial_t\boldsymbol{\hat{u}} + \boldsymbol{\hat{u}}.\nabla(\boldsymbol{\hat{u}}) &= -\nabla \hat{p} + \frac{1}{Re}\nabla^2 \boldsymbol{\hat{u}}, \;\;\mbox{and} \label{eq:mom}
    \\\nabla.\boldsymbol{\hat{u}} &= 0, \label{eq:cont}
\end{align}
where $Re$ is the Reynolds number, and $\boldsymbol{\hat{u}}, \hat{p}$ are velocity and pressure fields satisfying the boundary conditions as shown in the problem setup schematic in Fig.~\ref{fig:problemsetup}, respectively. On the moving boundary $\Gamma_{IB}$ especially, a no-slip boundary condition dependent on the plunging kinematics is to be satisfied. Here, the plunging kinematics is modeled as follows
\begin{align} 
	{y}({t}) &= h_0 \cos(\omega_h {t}), \;\; \mbox{and} \label{eq:kin1}
	\\\dot{{y}}({t}) &= -\omega_h h_0 \sin(\omega_h {t}), \label{eq:kin2}
\end{align}
where $t, h_0$ and $\omega_h$ are dimensional time, the plunging amplitude and frequency, respectively. The non-dimensional amplitude $h = h_0/c$ and reduced frequency $k = \omega_h c / U_{\infty}$ are defined aligning with literature~\cite{lewin2003modelling,khalid2018bifurcations}, where, $U_{\infty} = 1 $ is the free stream velocity and $c = 1$ is the chord length of the foil. 

In the present study, the example from \cite{sundar2023physicsinformed} with $Re= 500, k=2\pi$ and $h=0.16$ is chosen aligning with \cite{khalid2018bifurcations} for gradient-based analysis. The training data has been generated using a discrete forcing IBM-based unsteady flow solver~\cite{majumdar2020capturing} where, additional momentum forcing ($\boldsymbol{f}$) and source/sink ($q$) terms are added to Eqs.~(\ref{eq:mom}) and (\ref{eq:cont}), respectively to satisfy the no-slip boundary condition.  
 
\subsection{\textbf{Moving boundary enabled PINNs (MB-PINNs)}}
\begin{figure*}[!htbp]
	\centering
	\includegraphics[width=0.9\linewidth]{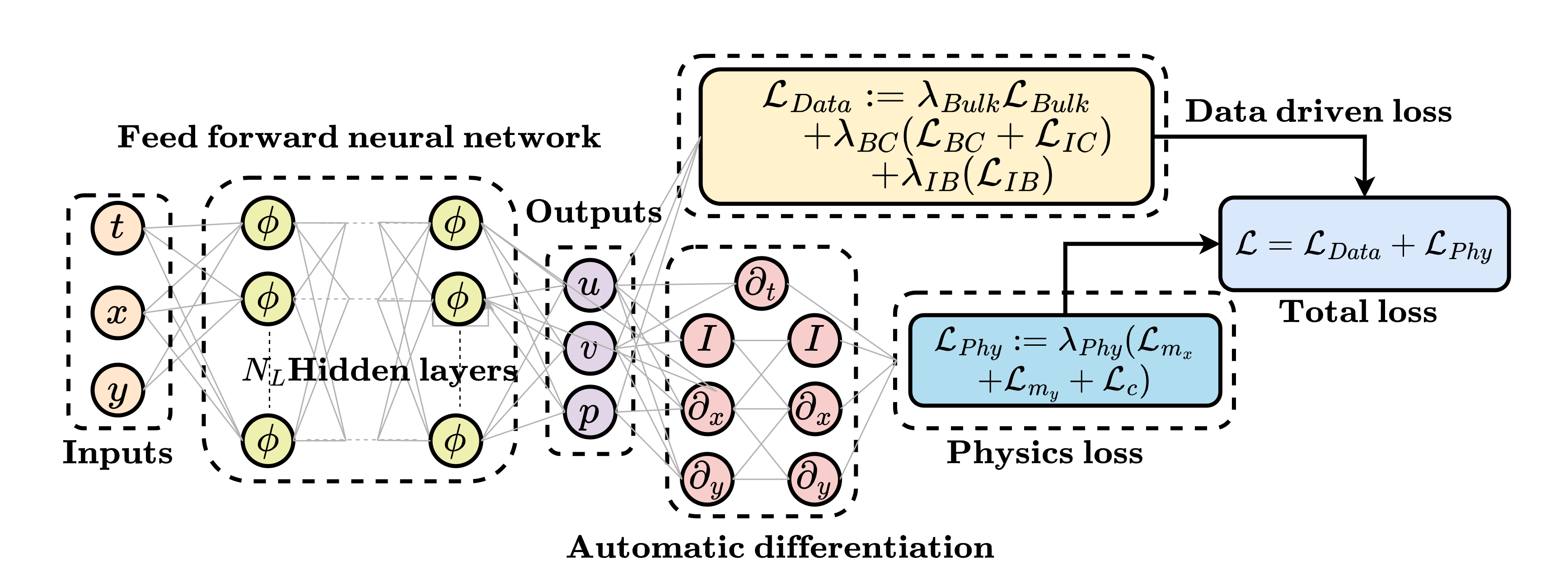}
	\caption{MB-PINN schematic (Image adapted from Sundar {\it et al.}~\cite{sundar2023physicsinformed}.)}
	\label{fig:mb-pinnmodifiedfmfp}
\end{figure*}

Inspired by the immersed boundary method, an immersed boundary aware framework using PINNs for surrogate modeling of unsteady flows past moving boundaries was explored by Sundar {\it et al.}~\cite{sundar2023physicsinformed}. A truncated spatial domain excluding solid region points $\Omega_f^r$ was chosen as shown in Fig.~\ref{fig:problemsetup} to train the moving boundary enabled PINN(MB-PINN) models in ~\cite{sundar2023physicsinformed}.
It was shown that the MB-PINN (see Fig.~\ref{fig:mb-pinnmodifiedfmfp}) was efficacious against solving velocity reconstruction and simultaneous pressure recovery given IBM data. A feed forward neural network with Swish activation as a backbone was used in ~\cite{sundar2023physicsinformed} to map the spatio-temporal coordinates to the corresponding velocity and pressure predictions. The loss function $\mathcal{L}$ can be written as follows 

\begin{align}
    \mathcal{L} &= \lambda_{Bulk}\mathcal{L}_{Bulk} + \lambda_{BC}\mathcal{L}_{BC} +\lambda_{IC}\mathcal{L}_{IC} \notag\\&+\lambda_{IB}\mathcal{L}_{IB} + \lambda_{Phy}\mathcal{L}_{Phy}.\label{eq:mbpinnloss}
\end{align}

Here, $\mathcal{L}_\#$ with the subscripts $\# = \{\text{Bulk, BC, IC, IB, Phy}\}$ denote the loss contributions from mean squared errors in interior bulk velocity data, boundary condition, initial conditions, no-slip boundary condition on the immersed boundary and the physics constraints, respectively and $\lambda_{\#}$ denotes the corresponding loss weights. Here, the boundary conditions include Dirichlet values at the inlet, upper and lower boundaries of the truncated computational domain $\Omega_f^r$ (see Fig.~\ref{fig:problemsetup}). 

Throughout the study, $\lambda_{Bulk} = \lambda_{BC} = \lambda_{IC} = \lambda_{IB} = 1,$ and the physics loss alone is relaxed.
In addition to physics loss relaxation, a vorticity cutoff based sampling was proposed in ~\cite{sundar2023physicsinformed}. Here, a vorticity cutoff $|\omega_z^*|,$ is chosen and all those fluid data points retained such that $|\omega_z|\geq|\omega_z^*|.$ 
Then based on a percentage sampling ratio $\displaystyle S_{\omega_z} = \frac {N^{sample}_{|\omega_z|<|\omega_z^*|}}{N_{|\omega_z|<|\omega_z^*|}} \times 100,$ the remaining fluid points are undersampled.

Considering $K_{MB}$ mini batches of the data, the network weights and biases together represented by $\theta$ are updated using the back propagated loss component gradients as follows 
 \begin{align}
   \boldsymbol{\theta}^{i+1} &= \boldsymbol{\theta}^{i} - \eta_i \frac{1}{N_{MB}}\sum_{i' = iN_{MB}+1}^{(i+1)N_{MB}}\bigg(\lambda_{Bulk}\nabla_{\boldsymbol{\theta}}\mathcal{L}_{Bulk}^{i'}\nonumber 
 \\&+ \lambda_{BC}\nabla_{\boldsymbol{\theta}}\mathcal{L}_{BC}^{i'}\; +\;  \lambda_{IB}\nabla_{\boldsymbol{\theta}}\mathcal{L}_{IB}^{i'} \nonumber\\&+\; \lambda_{IC}\nabla_{\boldsymbol{\theta}}\mathcal{L}_{IC}^{i'}\;+\;\lambda_{Phy}\nabla_{\boldsymbol{\theta}}\mathcal{L}_{Phy}^{i'}\bigg),
 \end{align}
 
where $\eta_i$ is the learning rate corresponding to $i^{th}$ iteration and each epoch corresponds to $K_{MB} = N_{train}/N_{MB}$ iterations. Here, the MB-PINN models were trained using the stochastic gradient descent based optimisation algorithm ADAM~\cite{kingma2014adam}. Unless otherwise specified, a mini-batch size of $N_{MB} = 1500$ was chosen with a learning rate step decay $lr = [1e-03, 53-04, 1e-04]$ where each learning rate step consisted of 5e05 training iterations. 

The loss components $\mathcal{L}_{IB}, \mathcal{L}_{BC}$ don't depend on the interior spatial coordinates, and $\mathcal{L}_{IC}$ considers a data snapshot only at $t/T=0.$ They play an important role in ensuring the hidden variables are recovered appropriately ~\cite{buhendwa2021inferring, lucor2022simple}. However, $\mathcal{L}_{Bulk}$ and $\mathcal{L}_{Phy}$ are computed from spatial coordinates interior to $\Omega_f^r.$ Given that spatially, there are zones where strong, weak or no vortices exist, it would thus be interesting to investigate the contributions from different spatial regions internally in the domain.

\subsection{\textbf{Zonal splitting of layer-wise gradients}}
\begin{figure}[!t]
	\centering
	\includegraphics[width=0.7\linewidth]{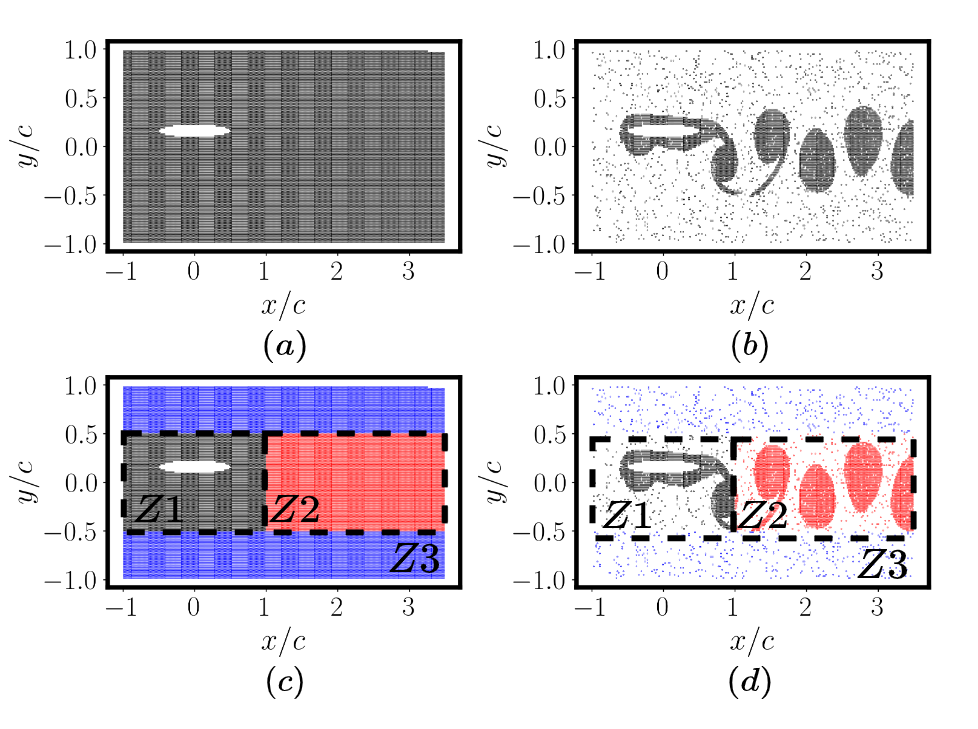}
	\caption{ Spatial grids of (a) CI and (b) CI-S5 data sets and the corresponding (c-d) zonal splitting of the grids into $\text{Z1}$- moving body, $\text{Z2}$ - wake and $\text{Z3}$ - outer zones, respectively}
	\label{fig:gridsampleszonalcomparison}
\end{figure}

While global loss weighting improves training and achieves global loss component balancing, that might not be the case locally. This could be due to varied contributions from each spatial region leading to different effective learning rates in each spatial region. To analyse this, once the MB-PINNs are trained or even at any intermediate learning stage, the layer-wise gradients can be computed overall for samples over the entire domain and for points sampled from select spatial zones. In the present study, three spatial zones are defined (see Fig.~ \ref{fig:gridsampleszonalcomparison}), namely, the moving body zone ( $Z1\subset \Omega^r_f$ such that  $-1< x <1 \text{ and } -0.5 < y < 0.5$ ), the wake zone ($Z2\subset \Omega^r_f$ such that  $1< x \text{ and } -0.5 < y < 0.5$ ), and the outer zone ( $Z3\subset \Omega^r_f$ such that $Z3 = \Omega^r_f - Z1\cup Z2 ).$ Note that the overall spatial domain is $\Omega_f^r$ which is $\Omega^r = [-1c,3.5c]\times[-1c,1c])$ minus the solid region points at any time $t\in[0,2].$ Also note that, zonal splitting is not carried out in temporal domain.

Given training velocity data $\boldsymbol{\hat{u}}$ at these spatial coordinates across all time, the components $\mathcal{L}_{Bulk}$ and $\mathcal{L}_{Phy}$ in Eq.~\ref{eq:mbpinnloss} can hence be split into contributions from respective spatial-zones as follows
\begin{align}
\mathcal{L}_{Bulk} &= \frac{1}{N_{MB}}\bigg(\sum_{i = 1}^{3}\bigg(\sum_{j = 1}^{N_{Zi}}\|\boldsymbol{u}^j- \boldsymbol{\hat{u}}^j\|^2\bigg)\bigg)
\\ &= \bigg(\sum_{i = 1}^{3}\bigg(\frac{N_{Zi}}{N_{MB}}\sum_{j = 1}^{N_{Zi}}\frac{\|\boldsymbol{u}^j - \boldsymbol{\hat{u}}^j\|^2}{N_{Zi}} \bigg)\bigg)
\\&= \bigg(\sum_{i=1}^{3}\bigg(p_{Zi}\sum_{j = 1}^{N_{Zi}}\frac{\|\boldsymbol{u}^j - \boldsymbol{\hat{u}}^j\|^2}{N_{Zi}}\bigg)\bigg)
\\&=\bigg(\sum_{i = 1}^{3}\bigg(p_{Zi} \mathcal{L}_{Bulk}^{Zi}\bigg)\bigg)
\end{align}
It is seen that $\mathcal{L}_{Bulk}$ is a weighted sum of zonal mean squared errors where $p_{Zi}$ for $i = 1,2,3$ is the proportion of sample points from zone $Zi$ with respect to the overall mini-batch sample. 
Similarly, $\mathcal{L}_{Phy}$ can be split into respective contributions from the spatial zones as follows
\begin{equation}
\mathcal{L}_{Phy} = \bigg(\sum_{i=1}^{3}\bigg(p_{Zi}\sum_{j = 1}^{N_{Zi}}\frac{\|\boldsymbol{r}(\boldsymbol{u}^j)\|}{N_{Zi}}\bigg)\bigg)
\end{equation}
The above expressions can be simplified as 
\begin{equation}
\mathcal{L}_*=\bigg(\sum_{i = 1}^{3}\bigg(p_{Zi} \mathcal{L}_*^{Zi}\bigg)\bigg)
\end{equation}
for $* = \{\text{Bulk, Phy}\}.$
Now the gradient vector $\nabla_{\theta}\mathcal{L}_*$ required to be computed at every training iteration to update the network parameters can be further expressed in terms of the zonal contributions as 
\begin{equation}
\nabla_{\theta}\mathcal{L}_* =\bigg(\sum_{i = 1}^{3}\bigg(p_{Zi} \nabla_{\theta}\mathcal{L}_{*}^{Zi}\bigg)\bigg).
\end{equation}

The above zonal splitting further indicates that in addition to loss component weighting, the data and collocation points sampling can modify the contributions from spatial zones or even spatio-temporal zones of interest. Hence, the zonal contributions can't be investigated independent of the sample proportions as the overall loss gradient is the weighted sum of the gradients obtained over each zone.

\subsection{\textbf{Metrics to diagnose zonal imbalances}}
Visualising and comparing the distribution of layerwise loss component gradients often sheds light on the extent of balance between the components~\cite{wang2020understanding} globally or locally. In addition, as seen previously, since gradient updates also depend on the proportion of zone specific coordinates with respect to the overall sample $p_{Z\#},$ this has to be weighed in while quantifying the relative contributions from each spatial zone. 
To quantify which zone drives the training for components $\mathcal{L}_*$ with $* = \{\text{Bulk, Phy}\},$ the following zonal proportion weighted relative gradient statistics based metrics are thus computed
\begin{align}
    \mu^{Zi}_{*} &= p_{Zi}\frac{\overline{|\nabla_{\theta}\mathcal{L}_{*}^{Zi}|}}{\overline{|\nabla_{\theta}\mathcal{L}_{*}|}} \label{eq:mu}
    \\\sigma^{Zi}_{*} &= p_{Zi}\frac{\text{std}\{|\nabla_{\theta}\mathcal{L}_{*}^{Zi}|\}}{\text{std}\{|\nabla_{\theta}\mathcal{L}_{*}|\}}.\label{eq:sigma}
\end{align}
Once MB-PINNs are trained, the mean and standard deviation of the gradient magnitudes $|\overline{\nabla_{\theta}\mathcal{L}_*}|$, $\text{std}\{|\nabla_{\theta}\mathcal{L}_*|\}$ are computed over the network parameters for a mini-batch sample from the entire spatio-temporal domain and also for samples from respective spatial zones. The metrics $\mu^{Zi}_{*}, \sigma^{Zi}_{*}$ for $i  = \{Z1, Z2, Z3\},$ and $* = \{\text{Bulk, Phy}\}$ are then computed. 

\section{\textbf{RESULTS AND DISCUSSION}}\label{sec3}
\begin{figure}[!t]
	\centering
	\includegraphics[width=0.6\linewidth]{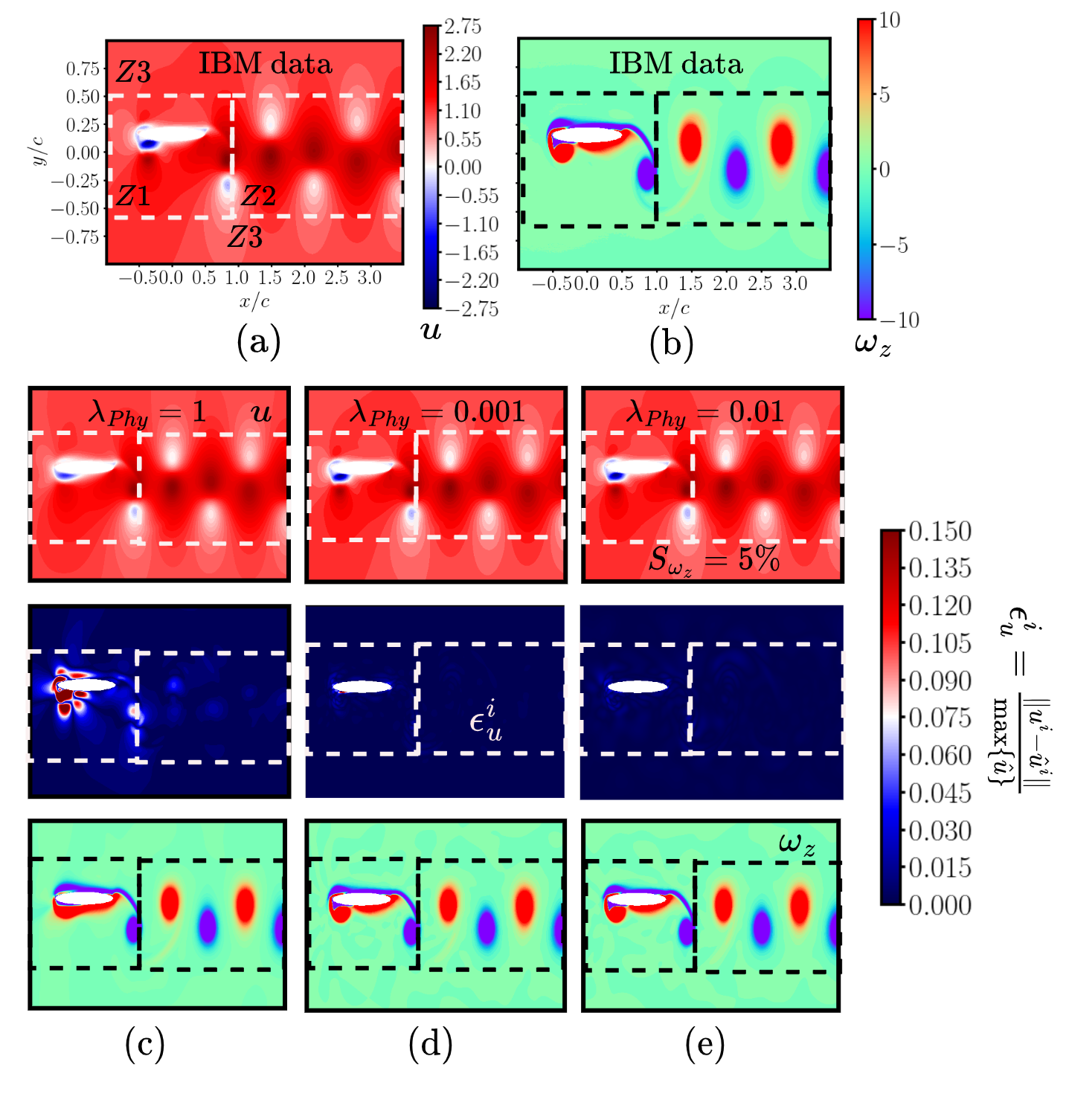}
	\caption{Comparison of (a-b) true and (c-e) MB-PINN predicted velocity $x$-component, corresponding point-wise maximum value normalised absolute errors and vorticity contours at $t/T = 1.0.$}
	\label{fig:uvpcontoursbest}
\end{figure}

In the present study, three MB-PINN test cases from ~\cite{sundar2023physicsinformed} are considered for analysis; a model without any physics loss relaxation or under sampling (case 1), a model with physics loss relaxation alone (case 2), and a model with both physics loss relaxation and vorticity cutoff based undersampling (case 3). The true and predicted velocity x-component, associated point-wise errors, and vorticity contours for these cases are presented in Fig.~\ref{fig:uvpcontoursbest} where the errors are prominent in zone $Z1$ and $Z2$ for case 1 but they almost vanish for cases 2 and 3 (see Figs.~\ref{fig:uvpcontoursbest}(c-e)). The accuracy details are also presented in Table ~\ref{tab:testcaseaccuracy}. To further understand which spatial zones in the flow contribute to the training and determine if there exist spatial imbalances, the loss component wise gradient distributions are analysed. The metrics $\mu_{*}^{Zi}$ and $\sigma_{*}^{Zi}$  for $i = 1,2,3$ and $* = \{\text{Bulk, Phy}\}$ 
 described in Eqs.~(\ref{eq:mu}) and (\ref{eq:sigma}) are presented in Table~\ref{tab:gradstats} for all the test cases discussed below. 

\begin{table}[t!]
	\centering
	\caption{ Training and testing data set resolution within the truncated domain considered in this study. Ref-IBM and Ref-ALE are the high resolution testing data-sets generated using a IBM and an ALE solver, respectively. Further details about the 'Ref-*' data-sets can be found in ~\cite{sundar2023physicsinformed}. }
	\begin{tabular}{cccccc}
		\hline
		\textbf{Data sets} & $N_x$ & $N_y$ & $N_t$& $\Delta t/T$ &($Nx\times Ny\times N_t$) \\
		\hline
		CI & 270 & 120 & 41 & 0.05 & \textbf{1.3284e06}\\
            CI-S5 & - & - & - & 0.05 & \textbf{2.648e05}\\
            Ref-IBM & 651 & 500 & 81 & 0.025 & 2.6365e07\\
		Ref-ALE & - & - & 81 & 0.025 & 4.05e06\\
		\hline
	\end{tabular}
	\label{tab:data-sets}
\end{table}

\begin{table}[h!]
	\centering
	\caption{Training data-sets and associated proportion $p_{Zi}$ of data points from each spatial zone $Zi$ with $i = 1,2,\mbox{and}\; 3$ for a mini-batch sample at$1e06$ iterations.}
	\begin{tabular}{ccccc}
		\hline
		Data set &  $N_{Bulk}$ & $ p_{Z1}$ & $p_{Z2}$ & $p_{Z3}$\\
		\hline
		CI & 1.2915e06 & 2.175e-01& 2.857e-01 & 4.967e-01\\
		CI-S5 & 2.6481e05 &2.936e-01  &5.841e-01 &1.223e-01\\
		\hline
	\end{tabular}
	\label{tab:vsdata-sets}
\end{table}

\begin{table}[t!]
	\centering
	\caption{Accuracy of MB-PINN for different relaxation coefficients evaluated on Ref-IBM and Ref-ALE datasets. Best performing models are highlighted in bold-faced fonts. }
	\gdef\rownumber{}
	\begin{tabular}[t]{ccccc}
		\hline
		\vspace{4pt}
		Test case & $\lambda_{fluid}$ & $S_{\omega_z}$ (in $\%$) & aMAE & arRMSE (in $\%$)\vspace{-2pt}\\
		\hline
		Case 1 & 1 & 100 &  1.37e-01 & 22.96 \\
		Case 2 & \textbf{0.001} & 100 	&\textbf{2.07e-02}	&\textbf{3.22}\\
		Case 3 &\textbf{ 0.01} &$\textbf{5}$	&\textbf{2.25e-02}&\textbf{3.22} \\
		\hline
	\end{tabular}
	\label{tab:testcaseaccuracy}
\end{table}

\subsection{\textbf{Case 1: Baseline ($\lambda_{Phy} = 1)$}}
\begin{figure}[!htbp]
    \centering
    \includegraphics[width = 0.75\linewidth]{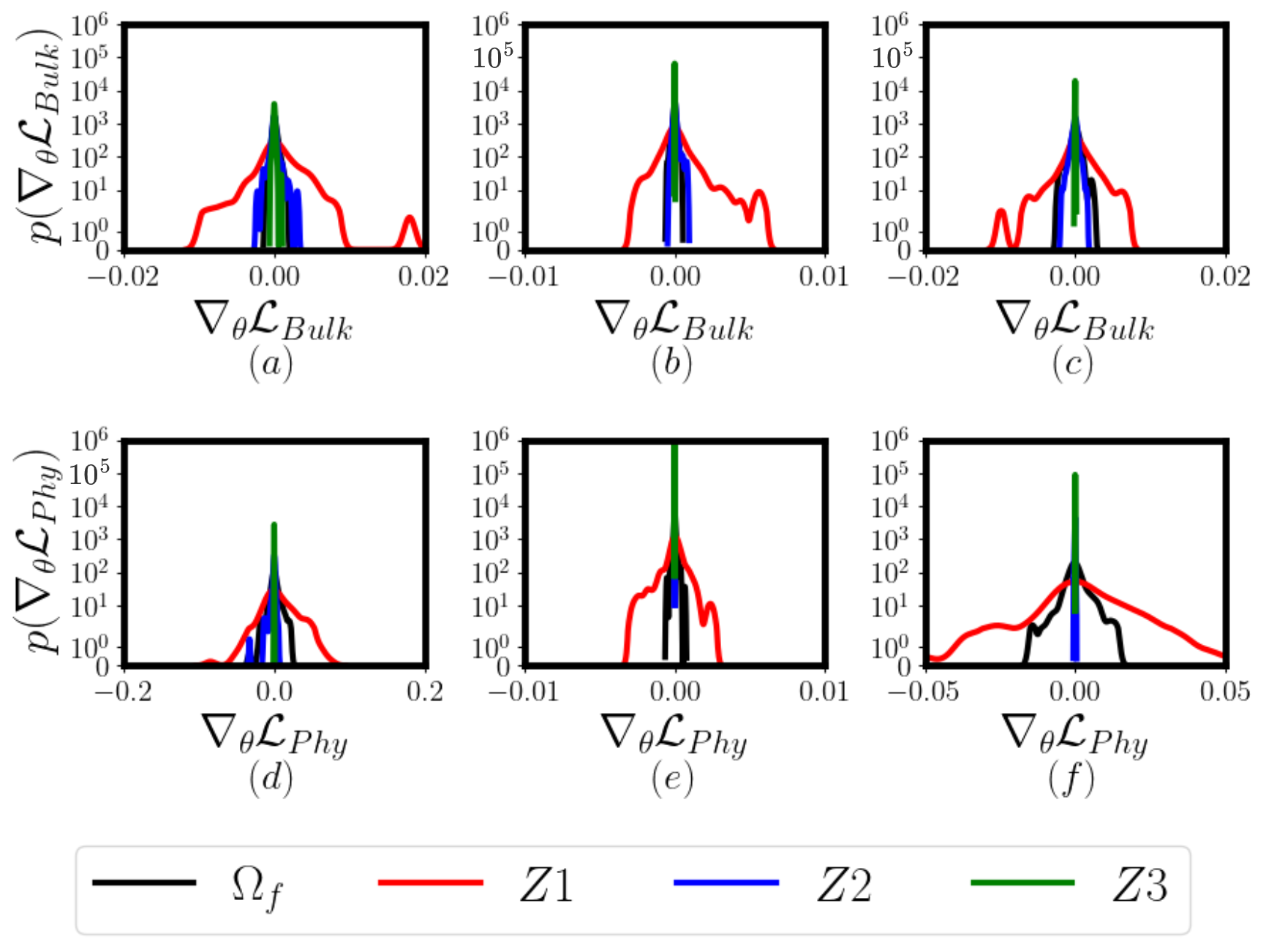}
    \caption{First layer $\nabla_{\theta}\mathcal{L}_{Bulk}$ (top row) and $\nabla_{\theta}\mathcal{L}_{Phy}$ (bottom row) distributions obtained after $1e06$ training iterations for (a,d) Case 1: $\lambda_{Phy}=1$, (b,e) Case 2: $\lambda_{Phy} = 0.001$ and $S_{\omega_z} = 100\%$ and (c,f) Case 3: $\lambda_{Phy} = 0.01$ and $S_{\omega_z} = 5\%.$}
    \label{fig:compare-first-zonal}
\end{figure}

It is observed that the first and last layer gradient distributions overall (see Figs.~\ref{fig:compare-overall}(a)and \ref{fig:compare-overall}(d)) indicate an imbalance in contributions more prominent in the first layer. The $\nabla_{\theta}{L}_{Phy}$ distributions across all zones are at least one order higher than that of $\nabla_{\theta}{L}_{Bulk}$ (see Figs.~\ref{fig:compare-first-zonal}(a)and \ref{fig:compare-first-zonal}(d) and Figs.~\ref{fig:compare-last-zonal}(a)and \ref{fig:compare-last-zonal}(d), respectively). Although the gradients have a higher $\mu$ and $\sigma$ (subscripts dropped for simplicity) in $Z1$ (see Table~\ref{tab:gradstats}) the gradients for the overall domain $\Omega_f$ still receive significant contributions from $Z2$ and $Z3$ (see Fig.~\ref{fig:compare-first-zonal}(a) and \ref{fig:compare-first-zonal}(d)) since $p_{Z1} < p_{Z2} < p_{Z3}.$    
\subsection{\textbf{Case 2: With physics loss relaxation ($\lambda_{Phy} = 0.001$)}}
\begin{figure}[!htbp]
    \centering
    \includegraphics[width = 0.75\linewidth]{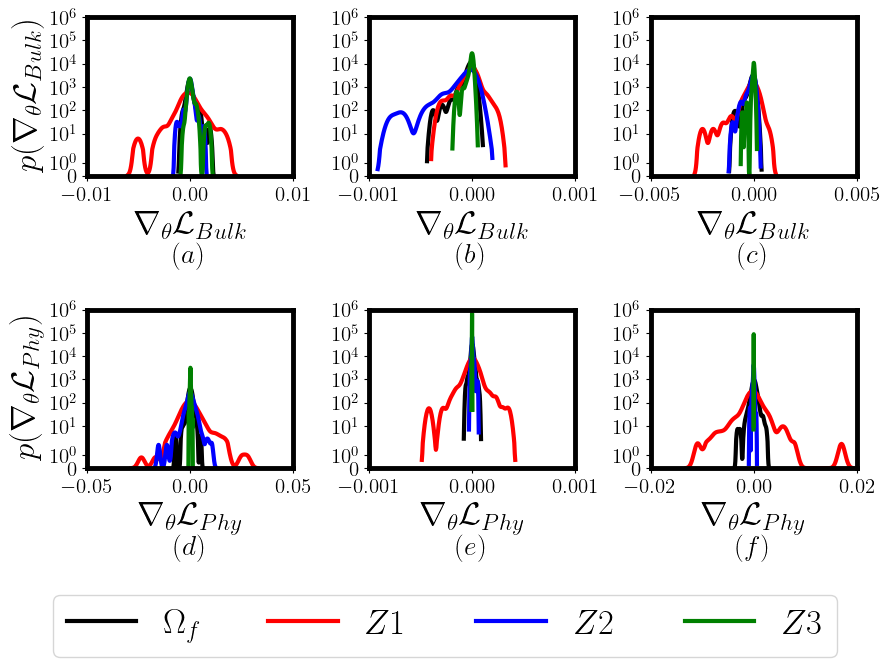}
    \caption{Comparison of last layer $\nabla_{\theta}\mathcal{L}_{Bulk}$ (top row) and $\nabla_{\theta}\mathcal{L}_{Phy}$ (bottom row) distributions obtained after $1e06$ training iterations for (a,d) Case 1: $\lambda_{Phy}=1$, (b,e) Case 2: $\lambda_{Phy} = 0.001$ and $S_{\omega_z} = 100\%$ and (c,f) Case 3: $\lambda_{Phy} = 0.01$ and $S_{\omega_z} = 5\%.$}
    \label{fig:compare-last-zonal}
\end{figure}

With global physics loss relaxation, the MB-PINN model trained on the CI dataset was reported to be optimal for $\lambda_{Phy} = 0.001.$ It is seen that the gradient magnitudes are much lower than that of the baseline case indicating some sort of vanishing gradient problem (see Figs.~\ref{fig:compare-overall}(b)and \ref{fig:compare-overall}(e)). Even though the gradient magnitudes are very low, the loss component gradients are relatively well balanced across all the zones unlike in case 1 (see Figs.~\ref{fig:compare-first-zonal}(b)and \ref{fig:compare-first-zonal}(e), and Figs.~\ref{fig:compare-last-zonal}(b)and \ref{fig:compare-last-zonal}(e), respectively). Here, as seen in the Table~\ref{tab:gradstats}, $\nabla_{\theta}\mathcal{L}_{Bulk}$ is dominated by contributions from $Z2,$ while $\nabla_{\theta}\mathcal{L}_{Phy}$ receives most of its contributions from $Z1.$ In spite of the good accuracy, it is seen that the gradients almost vanish in the last layer as opposed to the first layer. 
\subsection{\textbf{Case 3: With global physics loss relaxation and vorticity cutoff based under sampling ($\lambda_{Phy} = 0.01, S_{\omega_z} = 5\%$)}}
\begin{figure}[!htbp]
    \centering
\includegraphics[width = 0.75\linewidth]{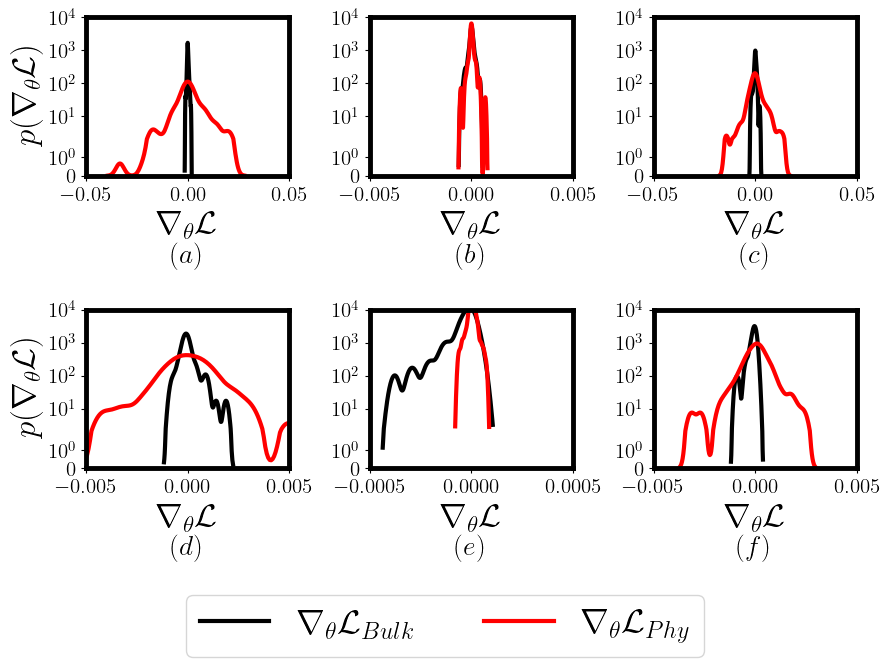}
    \caption{First layer (top row) and last layer (bottom row) $\nabla_{\theta}\mathcal{L}_{Bulk}$ and $\nabla_{\theta}\mathcal{L}_{Phy}$ distributions obtained after $1e06$ training iterations for (a,d) Case 1: $\lambda_{Phy}=1$, (b,e) Case 2: $\lambda_{Phy} = 0.001$ and $S_{\omega_z} = 100\%$ and (c,f) Case 3: $\lambda_{Phy} = 0.01$ and $S_{\omega_z} = 5\%.$}
    \label{fig:compare-overall}
\end{figure}

Combining the physics loss relaxation with a physics based vorticity cutoff sampling was identified as a useful strategy in \cite{sundar2023physicsinformed} to reduce the data requirement while still obtaining similar accuracy as in case 2 (see Table~\ref{tab:testcaseaccuracy}). It is observed that in case 3 the magnitudes of $\nabla_{\theta}\mathcal{L}_{Bulk}$ and $\nabla_{\theta}\mathcal{L}_{Phy}$ are higher and they don't vanish unlike in case 2 (see Figs.~\ref{fig:compare-overall}(c) and \ref{fig:compare-overall}(f)). In a way, since vorticity cutoff based under sampling retains more data points only in those regions with strong gradients, it is possible that this prevents the gradients from vanishing. Unlike Case 1 and Case 2, due to selective under sampling, $p_{Z3}< p_{Z1} <p_{Z2}$ which in turn affects the gradient back propagation. As a result, in Table~\ref{tab:gradstats} it is seen that $\nabla_{\theta}\mathcal{L}_{Bulk}$ and $\nabla_{\theta}\mathcal{L}_{Phy}$ are both dominated by contributions from $Z1$ followed by that from $Z2.$ The first and last layer gradient distributions as well indicate that across all zones, the gradients are balanced with $Z1$ driving the training as seen by the gradient distrbutions on $\Omega_f$ following $Z1.$ 

\begin{table}[t!]
\centering
	\caption{Zonal relative gradient statistics across test cases. Dominant zones are highlighted in bold.}
	\gdef\rownumber{}
\begin{tabular}{ccccc}
    \hline
    \textbf{Zone} &$\mu_{Bulk}^{Z\#}$ & $\sigma_{Bulk}^{Z\#}$ &$\mu_{Phy}^{Z\#}$& $\sigma_{Phy}^{Z\#}$\\
    \hline
    \multicolumn{5}{c}{Case 1: $\lambda_{Phy} = 1$}\\
    \hline
     \textbf{Z1} & \textbf{7.932e-01} & \textbf{7.466e-01}& \textbf{1.256}& \textbf{2.153}\\
     Z2 & 2.179e-01 & 2.669e-01 & 1.292e-01 & 1.368e-01\\
    Z3 & 2.292e-01 &2.459e-01 & 6.503e-02 & 5.801e-02\\
     \hline
     \multicolumn{5}{c}{ Case 2: $\lambda_{Phy} = 0.001$}\\
    \hline
     \textbf{Z1} & 3.207e-01& 4.769e-01 &\textbf{5.565e-01} & \textbf{5.608e-01}\\
     \textbf{Z2} &\textbf{7.552e-01} & \textbf{8.477e-01} & 6.906e-02& 5.608e-02\\
     Z3 &1.297e-01 & 1.676e-01 & 9.610e-3& 5.823e-03 \\
     \hline
     \multicolumn{5}{c}{Case 3: $\lambda_{Phy} = 0.01,$ $S_{\omega_z} = 5\%$}\\
    \hline
     \textbf{Z1} & \textbf{7.054e-01} &\textbf{1.194} & \textbf{4.929}& \textbf{7.555}\\
     Z2 & 4.872e-01 & 4.392e-01 &1.359e-01 & 1.073e-01\\
      Z3 & 7.959e-03 & 5.917e-03& 1.432e-03 & 7.506e-04\\
     \hline
\end{tabular}
\label{tab:gradstats}
\end{table}


\section{\textbf{CONCLUSIONS}}\label{sec4}
In this paper, to understand the training of MB-PINNs for flow past a plunging foil, a novel zonal splitting methodology and gradient statistics-based metrics were proposed to analyze spatial imbalances in loss component contributions. Three zones consisting of the moving body, the wake and the outer far-field were considered. Three test cases from ~\cite{sundar2023physicsinformed} were considered for analysis with and without physics loss relaxation and vorticity cutoff based under-sampling. The analysis confirmed the existence of zonal imbalances and also determined which zone dictated training. In the test case with the relaxation of the physics loss alone, the wake zone dominated the gradient updates coming from the loss of data. Whereas, the moving body zone dominated the updates from physics loss. However, for the test case with both physics loss relaxation and physics-based undersampling, the moving body zone dominated the gradient updates for both the loss components.
It was also observed that a vorticity cutoff based under-sampling alleviates the problem of vanishing gradients as opposed to the case of no under-sampling, further validating the importance of selective physics-based sampling. 
It is envisioned that quantifying the imbalanced contribution from spatial or even more generally input subdomains would enable the design of effective loss component weighting strategies. This approach can be extended effectively for other problems as well to evaluate imbalances in contributions from the input sub-domains of interest. Moreover, being agnostic to the model architecture, this approach can be applied to physics-based or even purely data-driven methods to determine which input subdomain plays a relatively active role in training.

\vspace{0.5cm}
\noindent
\textbf{ACKNOWLEDGEMENTS}\\
\noindent The authors would like to sincerely thank the P.G. Senapathy High-Performance Computing Centre of IIT Madras for providing the computing resources used in this study. \\

\bibliographystyle{ieeetr}
\bibliography{references.bib}

\end{document}